\documentstyle[prl,aps]{revtex}

\begin{document}
\draft
\title{Magnetotransport in the low carrier density ferromagnet EuB$_{6}$}                        
\author{S. S\"{u}llow$^{1,2}$, I. Prasad$^{1,}$\cite{pra}, S. Bogdanovich$^{1,}$\cite{bog}, M.C. Aronson$^{1}$, J.L. Sarrao$^{3,}$\cite{sar}, and Z. Fisk$^{3}$}
\address{$^{1}$Department of Physics, 2071 Randall Laboratory, University of Michigan, Ann Arbor, Michigan 48109--1120}
\address{$^{2}$Inst. f\"{u}r Metallphysik, TU Braunschweig, Mendelssohnstr. 3, 38106 Braunschweig, Germany}
\address{$^{3}$National High Magnetic Field Laboratory, 1800 E. Paul Dirac Drive, Florida State University, Tallahassee, Florida 32310}

\date{\today}
\maketitle

\begin{abstract}
We present a magnetotransport study of the low--carrier density ferromagnet EuB$_{6}$. This semimetallic compound, which undergoes two ferromagnetic transitions at $T_{l}$\,$=$\,15.3\,K and $T_{c}$\,$=$\,12.5\,K, exhibits close to $T_{l}$ a colossal magnetoresistivity (CMR). We quantitatively compare our data to recent theoretical work~\cite{maj}, which however fails to explain our observations. We attribute this disagreement with theory to the unique type of magnetic polaronic formation in EuB$_{6}$.
\end{abstract}


Recently, critical fluctuations have been proposed by Majumdar and Littlewood~\cite{maj} as mechanism causing the colossal magnetoresistivity (CMR) in a number of non--manganite materials, like in the pyrochlores or chalcogenide spinels~\cite{ramirez,shimikawa}. The authors of Ref.~\cite{maj} argued that, because in a ferromagnetic metal close to its critical point the dominant magnetic fluctuations are those with a wave vector $q$\,$\rightarrow$\,0, the contributions from these fluctuations to the resistivity $\rho$ should grow as Fermi number $k_{f}$ and the carrier density $n$ decrease. For sufficiently small $n$, like in ferromagnetic semimetals, a major part of the zero--field resistivity close to $T_{c}$ would be caused by magnetic fluctuations. Suppressing these in magnetic fields should generate the CMR in such materials. 

The resistivity of a low--carrier density system might also be affected by magnetic polarons. But magnetic polarons disappear ({\em i.e.} delocalize) if the magnetically correlated regions overlap, implying that critical magnetic scattering should dominate the resistivity for $k_{f}$\,$\xi(T)$\,$\gg$\,1 ($\xi(T)$: magnetic correlation length). In this regime of dominant critical scattering and in the clean limit, {\em i.e.} $k_{f}$\,$\lambda$\,$\gg$\,1, the low--field magnetoresistivity is quantitatively predicted to $(\rho (T, B) -\rho_{0} )/\rho_{0} = \Delta \rho /\rho = C (M/M_{sat})^{2}$, with $C$\,$\approx$\,($k_{f}$\,$\xi_{0}$)$^{-2}$ ($\lambda$: mean free path; $M$: magnetization; $\xi_{0}$: magnetic lattice spacing). Then, for a free electron gas a relationship between magnetoresistive coefficient and carrier density, $C$\,$\propto$\,($n^{2/3}$\,$\xi_{0}^{2}$)$^{-1}$, emerges as central result of Ref.~\cite{maj}, with a proportionality constant $\approx$\,1/38. 

To test this prediction we performed a detailed study of the magnetoresistive properties of the divalent cubic hexaboride EuB$_{6}$~\cite{fisk}. This semimetal, with a carrier density determined from quantum oscillation experiments of 8.8$\times$10$^{-3}$ electrons per unit cell, undergoes two ferromagnetic transitions at $T_{c}$\,=\,12.5\,K and $T_{l}$\,=\,15.3\,K~\cite{sullow}, derived here from the maxima in the temperature derivative of the resistivity, $d \rho/ d T$. The effective carrier masses are slightly smaller than the free electron mass, and the Fermi surface is almost spherical. Hence, a free electron model appropriately describes this compound. The zero--field resistivity is metallic, and in free electron approximation we find $k_{f}$\,$\lambda$\,$\gg$\,1 up to room temperature~\cite{lambda}. Further, with $\xi_{0}$\,$=$\,$\xi_{300 \rm K}$\,$=$\,4.185\,\AA , in Oernstein--Zernicke approximation, $\chi (0)$\,$\propto$\,$\xi^{2}(T)$, and with the experimentally determined dc--susceptibility $\chi_{0}$ from Ref.~\cite{sullow} the condition $k_{f}$\,$\xi(T)$\,$\gg$\,1 is fulfilled below 17\,K. Hence, EuB$_{6}$ fulfills all requirements of the model of Ref.~\cite{maj}.

Here, we present resistivity and magnetoresistivity measurements employing a standard 4--probe ac--technique on the crystal studied in Ref.~\cite{sullow}, with the current applied along the [100] and the field along the [010] of the crystalline unit cell. For a quantitative analysis of the magnetoresistivity we use the magnetization from Ref.~\cite{sullow}. 

In Fig.~\ref{fig:fig1}(a) and (b) we plot our raw data: the temperature ($T$) dependent resistivity $\rho$ of EuB$_{6}$ in fields $B$ up to 5\,T and the normalized magnetoresistivity $\Delta \rho (B) /\rho$ between 5.5 and 20\,K, corrected for demagnetization effects. The field dependence of $\rho$ reveals two different magnetoresistive regimes: For small $B$ a rapid decrease of $\rho (B)$ close to $T_{l}$ occurs, while hardly any effect on $\rho$ is observable below $\simeq$\,10\,K. The suppression of $\rho$ close to $T_{l}$, $\Delta \rho / \rho$\,$\approx$\,$-0.9$ in 2\,T, is comparable in size to other CMR compounds~\cite{ramirez,shimikawa}. In contrast, for large fields $\rho (B)$ increases with $B$, this in particular at low $T$. The positive magnetoresistivity represents the normal metallic contribution $\rho_{met}$ to $\rho (B)$. 

We extract the magnetic scattering contribution from the total magnetoresistivity by subtracting the metallic magnetoresistivity $\rho_{met}$. To do that we parametrize the high--field magnetoresistivity with $\rho_{met}$\,=\,$\rho_{0} + a B^{x}$, $x$\,$\approx$\,2 and derive the magnetic part $\rho_{mag}$\,$=$\,$\rho (B) - \rho_{met}$. The field dependence of $\rho_{met}$ thus established for the data at 13\,K is included in Fig.~\ref{fig:fig1}(b) as dashed line.

At any given temperature the minimum value of $\rho$ as function of $B$, $\rho_{min}$, constitutes an upper limit for the phonon contribution to $\rho$. We have included the values $\rho_{min}$ as function of $T$ in Fig.~\ref{fig:fig1}(a) as shaded area, illustrating that at and above $T_{c}$ phonons contribute less than 15\% to the zero--field resistivity. 

To examine the dependence of the magnetic magnetoresistivity $\Delta \rho_{mag} /\rho$ on the normalized magnetization $M/M_{sat}$ we plot the two quantities at different temperatures in Fig.~\ref{fig:fig2} in a log--log plot, with the magnetic field as implicit variable. In order to compare with the model of Ref.~\cite{maj}, which is valid only in the paramagnetic phase, we restrict our analysis to temperatures $T$\,$\geq$\,$T_{l}$. As is illustrated in Fig.~\ref{fig:fig2}, at these temperatures all data sets collapse on a universal curve. In particular, for $M/M_{sat}$\,$\leq$\,0.07 we find $\Delta \rho_{mag} /\rho$\,$=$\,$C$\,$(M/M_{sat})^{2}$, with 
$C$\,=\,75 (solid line).  

The value of $C$ is in striking contrast to the prediction of Ref.~\cite{maj}. With the carrier density $n$\,$=$\,1.2$\times$10$^{-4}$/\AA $^{3}$ for our crystal EuB$_{6}$ we compute $C$\,$\approx$\,(38\,$n^{2/3} \xi_{0}^{2}$)$^{-1}$\,=\,0.62 rather than the observed 75. More generally, following Ref.~\cite{maj} we plot $C$ vs. $n$\,$\xi_{0}^{3}$ for metallic ferromagnets and manganites in Fig.~\ref{fig:fig3}, together with the data for EuB$_{6}$ of our crystal and from previous works~\cite{guy,cooley}. In the plot we include the predicted values $C$\,$=$\,(38\,$n^{2/3} \xi_{0}^{2}$)$^{-1}$. The data for EuB$_{6}$ deviate by an order of magnitude from those of the other materials, emphasizing the vastly different magnetoresistive behavior of this compound. 

We believe that the unique type of magnetic polaron formation feature in EuB$_{6}$ causes the failure of the model of Ref.~\cite{maj} to account for the observed behavior. As we have proposed elsewhere~\cite{sul2}, at $T_{l}$ polaron metallization via magnetic polaron overlap leads to a drop of $\rho (T)$. The polaron metallization is accompanied by a filamentory type of ferromagnetic ordering, which arises from internal structure of the polarons. The bulk magnetic transition occurs at $T_{c}$. The field dependence of the resistivity close to $T_{l}$ then is mainly governed by the increase of the polaron size with magnetic field (rather than by the suppression of critical scattering, as suggested in Ref.~\cite{maj}), causing the metallization to occur at a higher temperature, and leading to the reduction of the resistivity at $T_{l}$ in magnetic fields.

Work at the University of Michigan was supported by the U.S. Department of Energy, Office of Basic Energy Sciences, under grant 94--ER--45526 and 97--ER--2753. Work at the TU Braunschweig was supported by the Deutsche Forschungsgemeinschaft DFG

\begin{figure}
\caption[resistance]{a.) The temperature ($T$) dependent resistivity $\rho$ of EuB$_{6}$ in zero field (--) and in fields $B$ of 0.021\,T\,($\Box$), 0.029\,T\,($\circ$), 0.05\,T\,($\bigtriangleup$), 0.076\,T\,($\bigtriangledown$), 0.1\,T\,($\diamond$), 0.2\,T\,(+), 0.5\,T\,($\times$), 1\,T\,($\bullet$) and 5\,T\,(- - -). The grey shaded area denotes an upper limit of the phonon contribution to $\rho$. b.) The normalized magnetoresistivity $(\rho(B) - \rho_{0})/\rho_{0}$\,$=$\,$\Delta \rho /\rho$ at 10\,K\,($\Box$), 13\,K\,($\circ$), 15.5\,K\,($-\bullet-$), 17.5\,K\,($\bigtriangledown$) and 20\,K\,($\diamond$). The dashed line (- - -) visualizes the field dependence of the metallic magnetoresistivity $\rho_{met} = \rho_{B=0} + a B^{x}$ for the data at 13\,K, which is used to extract the magnetic contribution $\rho_{mag} = \rho (B) - \rho_{met}$.} 
\label{fig:fig1}
\end{figure}

\begin{figure}
\caption[MversusR]{The magnetic magnetoresistivity $\Delta \rho_{mag} /\rho$ of EuB$_{6}$ vs. the normalized magnetization $M/M_{sat}$ at 15.5\,K\,($\Box$), 16\,K\,($\circ$), 17\,K\,($\bigtriangleup$), 18\,K\,($\bigtriangledown$), 19\,K\,($\diamond$) and 20\,K\,(+) in a log--log plot. The solid line represents $\Delta \rho_{mag} /\rho$\,$=$\,$75 \cdot (M/M_{sat})^{2}$.} 
\label{fig:fig2}
\end{figure}

\begin{figure}
\caption[pressure]{Experimental values $C$ vs. scaled carrier density $n$\,$\xi_{0}^{3}$ for metallic ferromagnets (solid symbols), manganites (open symbols) (from Ref.~\cite{maj}) and EuB$_{6}$: 1: this work; 2: Ref.~\cite{guy}, 3: Ref.~\cite{cooley}. The dashed line denotes the prediction $C$\,$=$\,(38\,$n^{2/3} \xi_{0}^{2}$)$^{-1}$, the solid line is a guide to the eye.} 
\label{fig:fig3}

\end{figure}


\begin{references}
\bibitem{lambda} We calculate at room temperature, with $\rho \approx 700 \mu \Omega$cm, in free electron approximation $k_{f} \lambda$\,=\,$\frac{(3 \pi^{2})^{2/3} \hbar}{e^{2} \rho n^{1/3}}$\,$\approx$\,10.
\end{references}
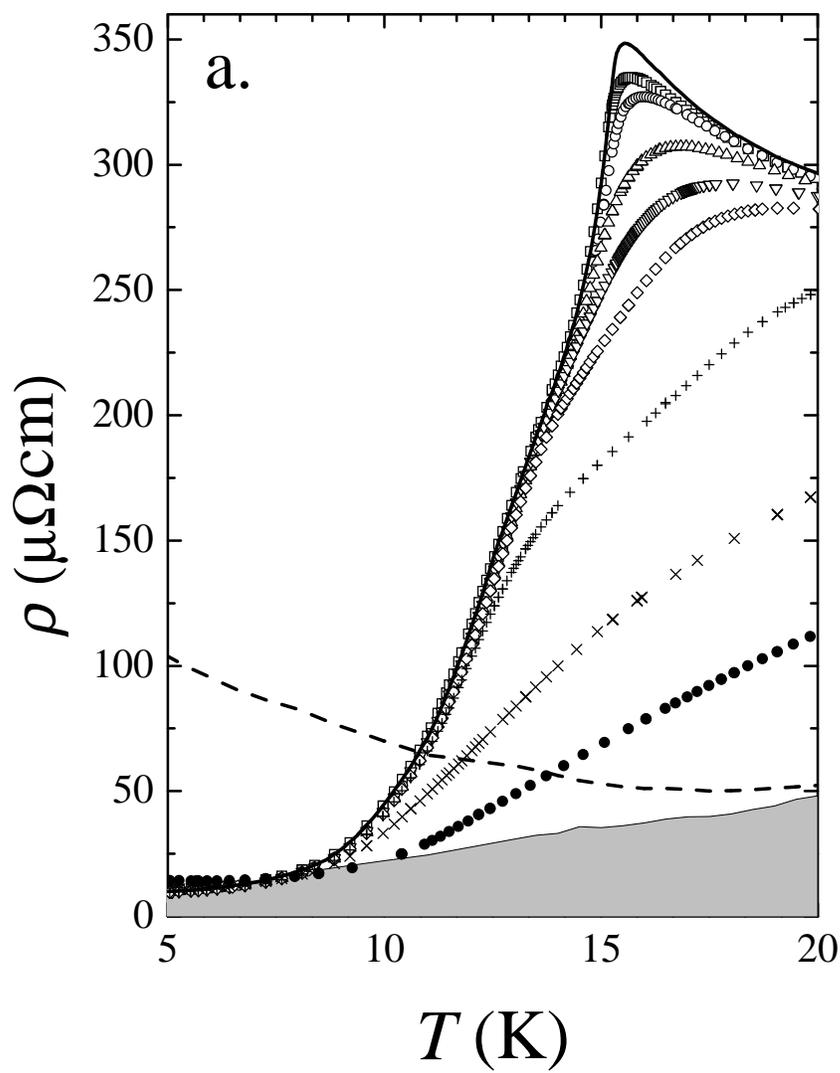
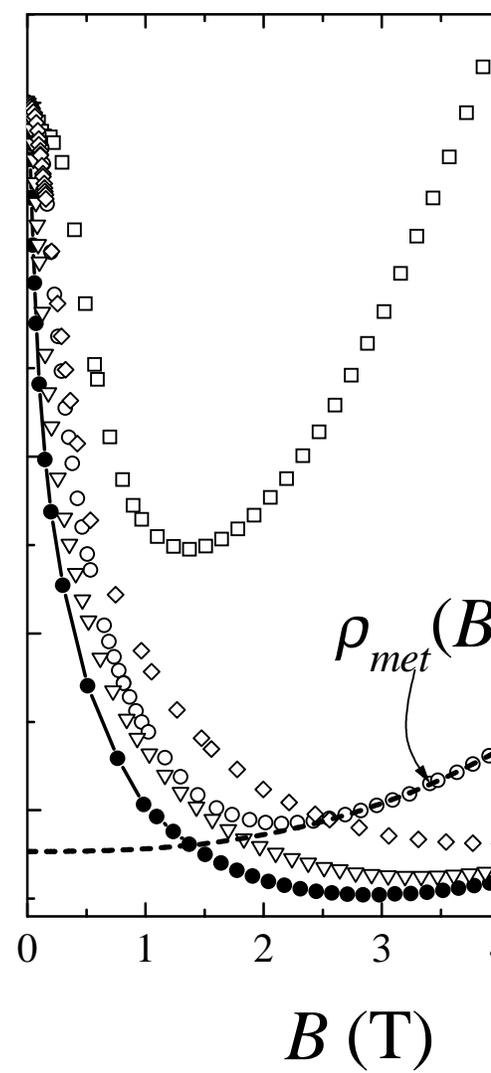
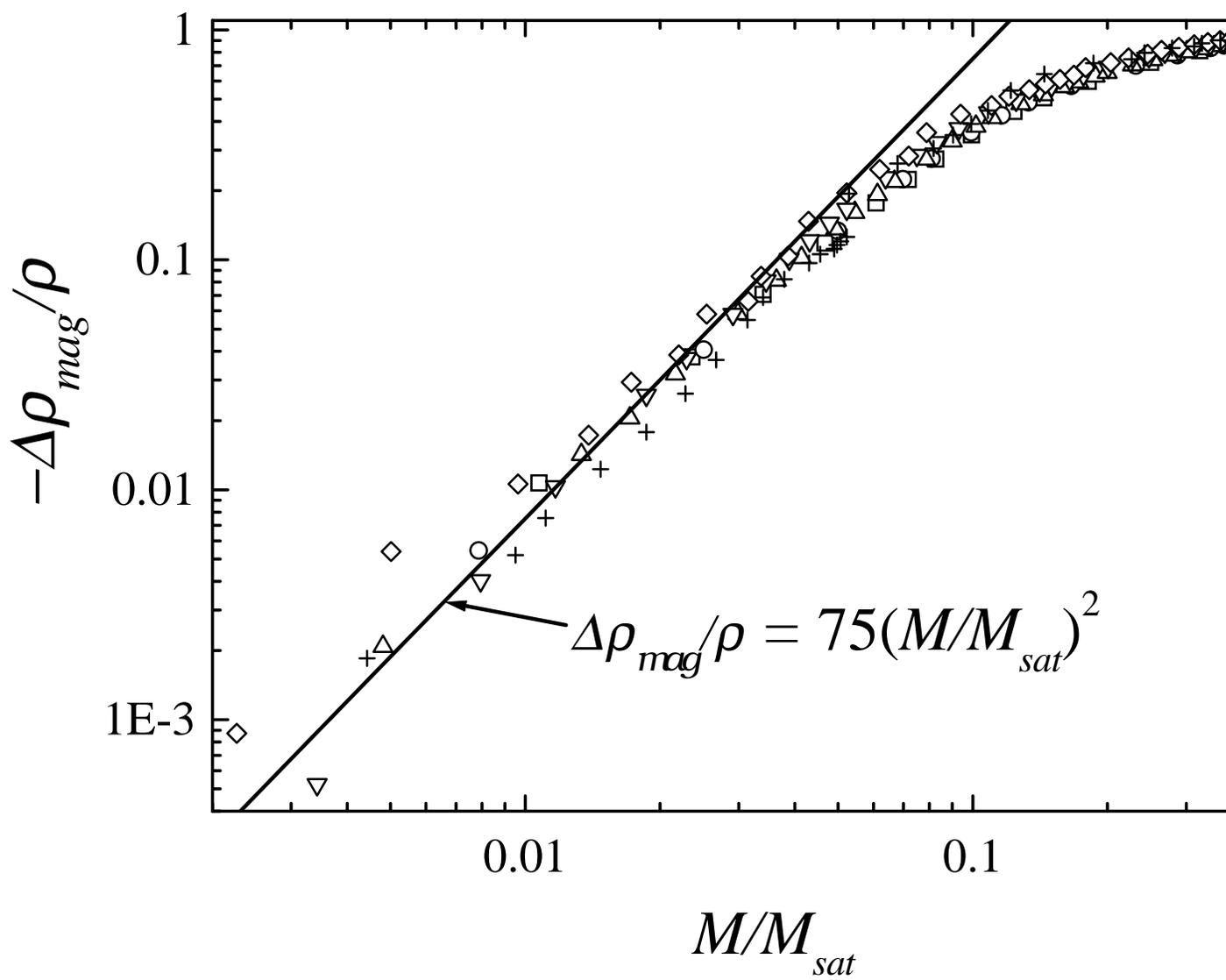
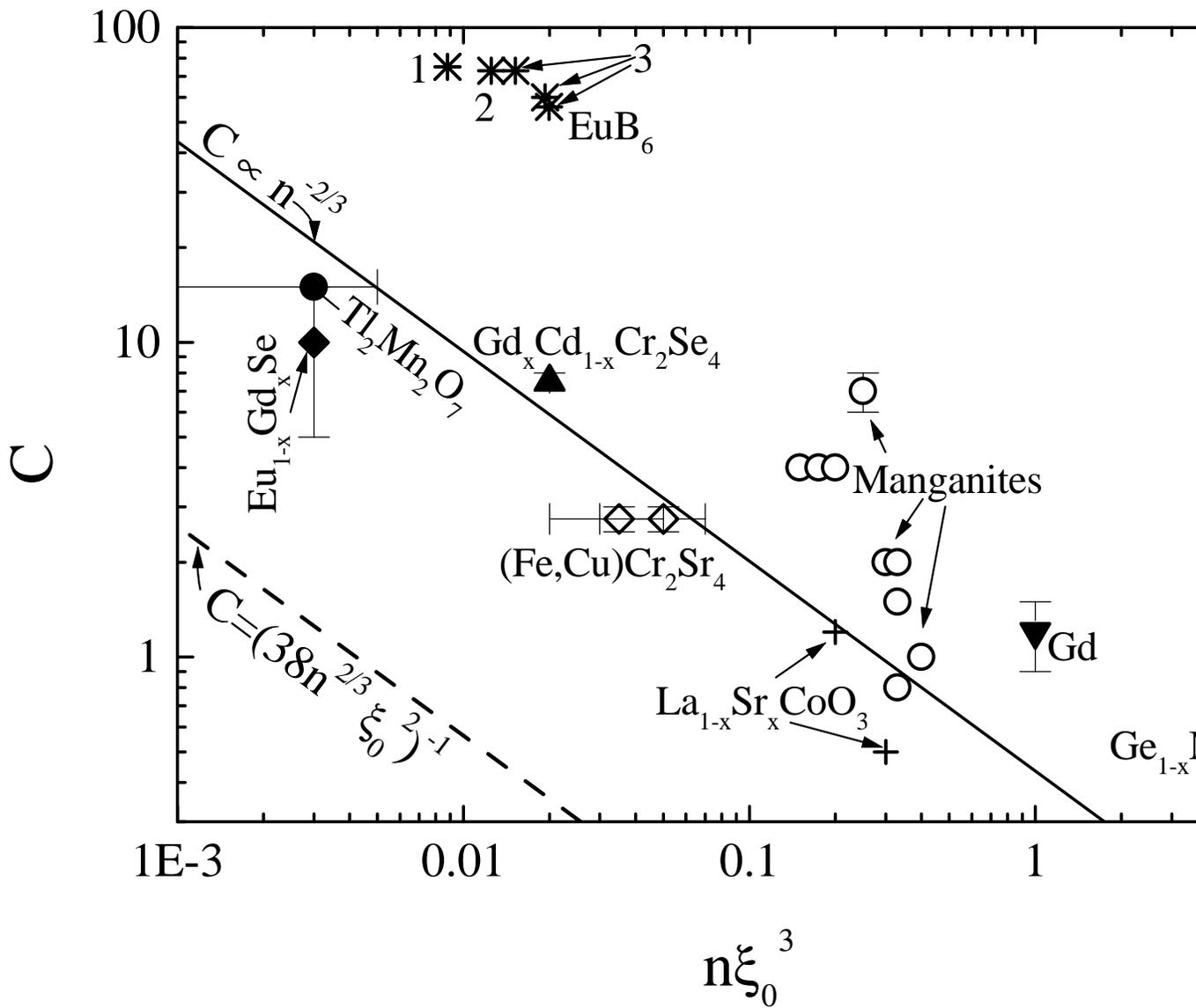
\end{document}